\documentclass[11pt]{amsart}
\usepackage{geometry}                % See geometry.pdf to learn the layout options. There are lots.
\usepackage{graphicx}
\usepackage{amssymb}
\usepackage{epstopdf}
\usepackage{longtable}
\usepackage{psfrag}
\usepackage{overpic}
\usepackage{color}
\usepackage{caption2}
\usepackage{subfigure}

\title{Magnetic fields induced by a mechanical singularity in a magnetoelastic half plane and their applications to the seismicities in the crust of the Chinese continent}

\author{Rong Qiang Wei}
\address{College of Earth Sciences, University of Chinese Academy of Sciences, Beijing, PRC, 100049}
\address{Key Laboratory of Computational Geodynamics, Chinese Academy of Sciences, Beijing, 100049, China}
\email{wrq1973@ucas.ac.cn}
\date{}
\begin{document}
\maketitle
\begin{abstract}
  The interaction between the magnetic field and the elastic deformation field in the crust is studied in a simplified way. The magnetic fields generated by the line singularities in a magnetized elastic half-plane are investigated. Using the general solution and Fourier transform technique, the exact solutions for the generated magnetic inductions due to various cases are obtained in a closed form. The results show that the line concentrated force will induced a perturbed magnetic field; The induced magnetic field will indicate the line concentrated force in reverse. The distribution of the vertical component of the magnetic induction caused by the line mechanical singularities is simpler than that of the horizontal component, and it is zero at the origin when the applied magnetic field and the line concentrated force satisfies some conditions. This result is applied to locate the epicenters of the earthquakes and historical earthquakes occurred in the Chinese continental crust. Results show that more than 80\% epicenters are at or near the zero-contours of the vertical component of the magnetic induction observed from satellite. These regions of zero-contours, especially those in active tectonic zones, could be the possible seismogenic zones in the future. The zero-contours of the vertical component of the magnetic induction from satellite could be as geophysical constraints to the risk information on the shallow seismicities, or they can be used as an early monitor for the shallow seismicities in the continental crust, or an auxiliary sign in the determination of the great historical earthquake.
  
 \end{abstract}

{\hspace{2.2em}\small Keywords:}

{\hspace{2.2em}\tiny Magnetic fields  mechanical singularity magnetoelasticity seismicity}

\section{Introduction}

The earth is a complex and multi-physical system. It is important to study the multi-physic fields and their coupled effects for understanding the properties of this complicated system. Here such a coupled effect of two physical fields--magnetoelasticity, which describing the interaction between the magnetic field and elastic deformation field in the crust, will be studied. 

The phenomenon of magnetoelasticity has received considerable attention in technology because of its applications in the non-destructive test (eg., Doubov, 1997). A general theory of magnetoelasticity had been developed by the approach of phenomenology since 1960’s (eg., Tiersten, 1964; Brown, 1966). However, such a general theory of magnetoelasticity is nonlinear and rather complicated. A linear magnetoelastical theory was developed by Pao and Yeh (1973), which has been used to investigate the magnetoelastic buckling of an isotropic plate in a uniform magnetic induction (Pao and Yeh, 1973), or the intensification factors of cracks in ferromagnetic solid (eg., Shindo, 1977; Lin and Yeh, 2002; Yeh and Ren, 2010). Based on the linear theory of magnetoelasticity above, Yeh (1989) studied the magnetic inductions generated by the mechanical singularities in a magnetized elastic half plane, and Huang and Wang (1995) presented a relatively general solution in a elastic magnetized half-space. In these studies, the direction of the applied magnetic induction is normal to the boundary of the half-plane (space), while the force is either parallel or normal to the boundary. In the similar work by Yeh and Ren (2010) which was performed in a magnetized anisotropic half plane, the direction of the applied magnetic induction varied but still the force is either parallel or normal to the boundary of the half plane. Besides, some modified models were presented to study the complicated magnetoelastic phenomenon (eg., Maugin, 1979; Zhou and Zheng, 1996; Qin et al., 2006).

   However, there are few applications of magnetoelastic in geoscience. Yeh (1987) investigated the induced magnetic potential generated by a tension fault in an infinite magnetized elastic plane. Wei and Yu (2012) studied the induced magnetic fields generated by the mechanical singularities in a magnetized elastic half plane, and used the related results to constrain the distribution of the epicenters for earthquakes in the continental crust of China. Similar to previous authors, the direction of the applied magnetic induction and the force are special.

   In this paper, the induced magnetic fields generated by the mechanical singularities in a magnetized elastic half plane will be studied, in which the direction of the applied magnetic induction and the force are random. The solutions are expressed in an analytical form and numerical results for several examples are presented. The application in the seismicities in the continental crust is discussed.

\section{Magnetoelastic problem of a magnetized elastic half plane}

\begin{figure}[htb]
\setlength{\belowcaptionskip}{0pt}
\centering
\begin{overpic}[scale=0.6]{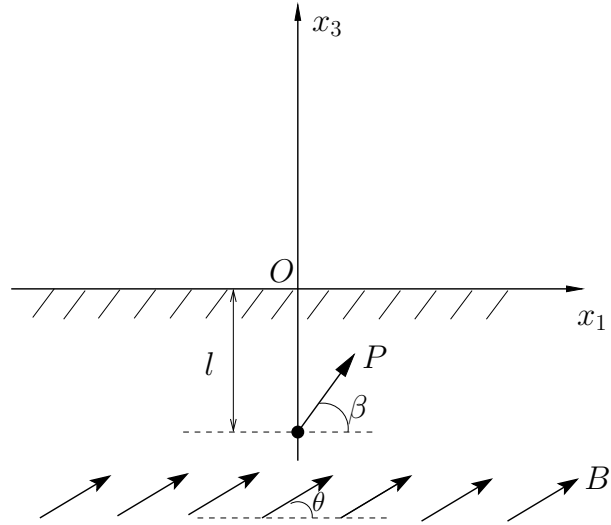}
\end{overpic}
\renewcommand{\figurename}{Fig.}
\caption{The model to be studied. A line mechanical singularity ${f_i}{{\bf{e}}_i}$ at $(0,{\rm{ }}-l)$ in the interior of a magnetized elastic half plane. The applied magnetic induction ${B_i}{{\bf{e}}_i}{\rm{ }}(i = 1,{\rm{ }}3)$  is distributed in whole plane uniformaly.}
\label{fig1}
\end{figure}

Although there are some works on 3D magnetoelastic problem (eg., Huang and Wang, 1995), only a 2D problem will be taken into account here for simplicity. Consider a magnetized elastic half plane occupying the region $- \infty  < {x_1} < \infty ,{x_3} \le 0$ as shown in Figure. \ref{fig1}. The applied magnetic induction ${B_i}{{\bf{e}}_i}{\rm{ }}(i = 1,{\rm{ }}3)$ is distributed in the whole plane uniformly. The induced magnetic field $b_i^ +$ by a line mechanical singularity ${f_i}{{\bf{e}}_i}$ at $(0,{\rm{ }} - l)$  is to be investigated based on the linear theory of magnetoelasticity introduced above for simplicity, although the general theory of magnetoelasticity is nonlinear and rather complicated (eg., Tiersten, 1964; Brown, 1966; Dorfmann and Ogden, 2014).

Mainly refer to Yeh (1989)，Huang and Wang (1995), and Yeh and Ren (2010), the corresponding differential equations and boundary conditions are as follows,

\begin{equation}\label{eq1}
\left\{ {\begin{array}{*{20}{l}}
\phi _{,ii}^+  = 0\\
{\phi _{,ii}} = 0\\
{u_{i,jj}} + \frac{1}{{1 - 2\nu }}{u_{j,ij}} + \frac{{2{\mu _0}\chi }}{G}{{\overline H }_j}{\phi _{,ij}} + \frac{{{f_i}}}{G} = 0
\end{array}} \right.
\end{equation}

At $x_3=0$,

\begin{equation}\label{eq2}
\left\{ {\begin{array}{*{20}{l}}
\phi _{,1}^ +  - {\phi _{,1}} =  - \chi {{\overline H }_3}{u_{3,1}}\\
\phi _{,3}^ +  - {\mu _r}{\phi _{,3}} =  - \chi {{\overline H }_1}{u_{3,1}}\\
{u_{1,3}} + {u_{3,1}} =  - \frac{{\chi {\mu _0}}}{G}({{\overline H }_1}{\phi _{,3}} + {{\overline H }_3}{\phi _{,1}})\\
\nu {u_{1,1}} + (1 - \nu ){u_{3,3}} = \frac{{(1 - 2\nu )\chi (\chi  - 2){\mu _0}}}{{2G}}{{\overline H }_3}{\phi _{,3}}\\
\overline {{H_1}}  = \frac{{{B_1}}}{{{\mu _0}}},{\rm{ }}\overline {{H_3}}  = \frac{{{B_3}}}{{{\mu _0}{\mu _r}}}
\end{array}} \right.
\end{equation}

where ${\phi ^ + }(b_i^ +  = {\mu _0}\phi _{,i}^ + )$ and $\phi$ are magnetic potential above or below the surface induced by the concentrated force which are finite when ${x_3} =  + \infty$ and ${x_3} =  - \infty$, respectively. $u_i$ are displacements,  $\overline{H}$ are magnetic intensity in the rigid-body state, and $f_i$ are body forces, respectively. $\mu_0, \mu_r$  are the magnetic permeability in vacuum and relative magnetic permeability, respectively. $\chi$ is the magnetic susceptibility. $\nu, G$ are elastic constants. $i,j=1,3$.  

\section{Induced magnetic field $b_i^ +$  by a line mechanical singularity}\label{sec3}

For the third equation(s) in Eq. (\ref{eq1}), the displacement can be discomposed into two parts as, ${u_i} = u_i^e + u_i^c$ , where $u_i^e$ is the special solution corresponding to the body force $f_i$, and $u_i^c$ the general solution of the corresponding homogeneous equations. In the case here, $u_i^e$ should satisfies the following conditions because of the zero-stress at the surface ${\sigma _{11}} = {\sigma _{33}} = {\sigma _{13}} = 0{\rm{  (}}{x_3} = 0{\rm{)}}$:

\begin{equation}\label{eq3}
\left\{ {\begin{array}{*{20}{l}}
{u_{1,3}^e + u_{3,1}^e = 0}&{{x_3} = 0}\\
{\nu u_{1,1}^e + (1 - \nu )u_{3,3}^e = 0}&{{x_3} = 0}
\end{array}} \right.
\end{equation}

so,

\begin{equation}\label{eq4}
\left\{ {\begin{array}{*{20}{l}}
{u_{1,3}} + {u_{3,1}} = u_{1,3}^c + u_{3,1}^c{\rm{ + }}u_{1,3}^e + u_{3,1}^e{\rm{ = }}u_{1,3}^c + u_{3,1}^c\\
\nu {u_{1,1}} + (1 - \nu ){u_{3,3}} = \nu u_{1,1}^c + (1 - \nu )u_{3,3}^c{\rm{ + }}\nu u_{1,1}^e + (1 - \nu )u_{3,3}^e{\rm{ = }}\nu u_{1,1}^c + (1 - \nu )u_{3,3}^c
\end{array}} \right.
\end{equation}

\begin{equation}\label{eq5}
\left\{ {\begin{array}{*{20}{l}}
{\phi _{,1}^ +  - {\phi _{,1}} =  - \frac{{\chi {B_3}}}{{{\mu _0}{\mu _r}}}u_{3,1}^c - \frac{{\chi {B_3}}}{{{\mu _0}{\mu _r}}}u_{3,1}^e}\\
{\phi _{,3}^ +  - {\mu _r}{\phi _{,3}} =  - \frac{{\chi {B_1}}}{{{\mu _0}}}u_{3,1}^c - \frac{{\chi {B_1}}}{{{\mu _0}}}u_{3,1}^e}
\end{array}} \right.
\end{equation}

It can be found from Eq. (\ref{eq4}) and (\ref{eq5}) that only $u_3^e$ is needed from the boundary conditions (\ref{eq2}) because of Eq. (\ref{eq3}). We obtained $u_3^e$  by the method of complex function analogous to Wang (2002) as follows,

\begin{equation*}
u_3^e = \frac{P}{{8\pi G(1 - \nu )}}\left\{ \begin{array}{l}
\sin \beta \{  - (3 - 4\nu )\log \hat r - [4(1 - \nu )(1 - 2\nu ) + 1]\log r\\
 + \frac{{(3 - 4\nu ){{({x_3} - l)}^2} + 2l{x_3}}}{{{r^2}}} + \frac{{{{({x_3} + l)}^2}}}{{{{\hat r}^2}}} - \frac{{4l{x_3}{{({x_3} - l)}^2}}}{{{r^4}}}\} \\
 + \cos \beta \{  - [4(1 - \nu )(1 - 2\nu )]\arctan \frac{{{x_1}}}{{{x_3} - l}} + \frac{{(3 - 4\nu ){x_1}({x_3} + l)}}{{{r^2}}}\\
\frac{{{x_1}({x_3} + l)}}{{{{\hat r}^2}}} + \frac{{4{x_1}{x_3}l({x_3} - l)}}{{{r^4}}}\} 
\end{array} \right\}
\end{equation*}

where $r = \sqrt {x_1^2 + {{({x_3} - l)}^2}} ,{\rm{ }}\hat r = \sqrt {x_1^2 + {{({x_3} + l)}^2}}$.

At $x_3=0$,

\begin{equation}\label{eq6}
u_{3,1}^e\left| {_{{x_3} = 0}} \right. = \frac{P}{{2\pi G}}\{ [(1 - 2\nu )\frac{l}{{x_1^2 + {l^2}}} + \frac{{{l^3} - lx_1^2}}{{{{(x_1^2 + {l^2})}^2}}}]\cos \beta  - [2(1 - \nu )\frac{{{x_1}}}{{x_1^2 + {l^2}}} + \frac{{2{l^2}{x_1}}}{{{{(x_1^2 + {l^2})}^2}}}]\sin \beta \}
\end{equation}

On the other hands,  $u_i^c$ can be assumed as follows analogous to Huang and Wang (1995),

\begin{equation*}
u_j^c =  - \frac{{2(1 - 2\nu ){\mu _0}\chi }}{G}\overline {{H_j}} \phi  + 4(1 - \nu ){P_j} - {({P_0} + {x_i}{P_i})_{,j}}
\end{equation*}

where ${P_{0,ii}} = {P_{i,jj}} = 0$ are harmonic functions to be determined later. Here according to our problem we assume that $P_1 = P_2 = 0$ and one can obtain:

\begin{equation}\label{eq7}
\left\{ {\begin{array}{*{20}{l}}
{u_1^c =  - \frac{{2(1 - 2\nu )}}{G}\chi {B_1}\phi  - {P_{0,1}} - {x_3}{P_{3,1}}}\\
{u_3^c =  - \frac{{2(1 - 2\nu )}}{{G{\mu _r}}}\chi {B_3}\phi  - {P_{0,3}} - {x_3}{P_{3,3}} + (3 - 4\nu ){P_3}}
\end{array}} \right.
\end{equation}

Submitting (\ref{eq7}) into (\ref{eq5}) and (\ref{eq4}), the boundary conditions now are as the following,

\begin{equation}\label{eq8}
\left\{ {\begin{array}{*{20}{l}}
{\phi _{,1}^ +  - [1 + \frac{{2(1 - 2\nu )}}{{G{\mu _0}\mu _r^2}}{\chi ^2}B_3^2]{\phi _{,1}} - \frac{{\chi {B_3}}}{{{\mu _0}{\mu _r}}}{P_{0,31}} + (3 - 4\nu )\frac{{\chi {B_3}}}{{{\mu _0}{\mu _r}}}{P_{3,1}}{\rm{ + }}\frac{{\chi {B_3}}}{{{\mu _0}{\mu _r}}}u_{3,1}^e = 0{\rm{     }}}\\
{\phi _{,3}^ +  - {\mu _r}{\phi _{,3}} - \frac{{2(1 - 2\nu )}}{{G{\mu _0}{\mu _r}}}{\chi ^2}{B_1}{B_3}{\phi _{,1}} - \frac{{\chi {B_1}}}{{{\mu _0}}}{P_{0,31}} + (3 - 4\nu )\frac{{\chi {B_1}}}{{{\mu _0}}}{P_{3,1}} + \frac{{\chi {B_1}}}{{{\mu _0}}}u_{3,1}^e = 0}\\
{\frac{\chi }{G}{B_1}(4\nu  - 1){\phi _{,3}} + \frac{\chi }{G}\frac{{{B_3}}}{{{\mu _r}}}(4\nu  - 1){\phi _{,1}} - 2{P_{0,31}} + 2(1 - 2\nu ){P_{3,1}} = 0{\rm{                          }}}\\
{ - \frac{{2\nu }}{G}\chi {B_1}{\phi _{,1}} - \frac{{\chi {B_3}}}{{G{\mu _r}}}[2(1 - \nu ) + \frac{{\chi  - 2}}{2}]{\phi _{,3}} - {P_{0,33}} + 2(1 - \nu ){P_{3,3}} = 0{\rm{                      }}}
\end{array}} \right.
\end{equation}

Hence our problem is reduced to that of solving 4 harmonic functions, i.e., ${\phi ^ + },\phi ,{P_0},{P_3}$. They can be determined by firstly employing Fourier transform with respect to $x_1$, then being substituted into boundary conditions Eq.(\ref{eq8}) to obtain the corresponding expressions, and finally employing Fourier inversion transform to these expressions. 

We take ${\phi ^ + }$ as an example. Employing Fourier transform to $\phi _{,ii}^ +  = 0$ with respect to $x_1$, we can obtain:

\begin{equation*}
\frac{{{{\rm{d}}^2}}}{{{\rm{d}}x_3^2}}A(\xi ,{x_3}) - {\xi ^2}A(\xi ,{x_3}) = 0
\end{equation*}

Because at $x_3= \infty$, $\phi^+$ is finite, $A(\xi ,{x_3}) = A(\xi )\exp ( - \left| \xi  \right|{x_3})$. Hence,

\begin{equation}\label{eq9}
{\phi ^ + } = \frac{1}{{2\pi }}\int_{ - \infty }^{ + \infty } {A(\xi )\exp ( - \left| \xi  \right|{x_3} + i\xi {x_1}){\rm{d}}\xi } 
\end{equation}

At $x_3=0$

\begin{equation}\label{eq10}
\begin{array}{l}
\phi _{,1}^ +  = \frac{1}{{2\pi }}\int_{ - \infty }^{ + \infty } {A(\xi )\exp (i\xi {x_1})(i\xi ){\rm{d}}\xi } \\
\phi _{,3}^ +  = \frac{1}{{2\pi }}\int_{ - \infty }^{ + \infty } {A(\xi )\exp (i\xi {x_1})( - \left| \xi  \right|){\rm{d}}\xi } 
\end{array}
\end{equation}

Similarly, the ${\phi _{,1}},{\phi _{,3}},{P_{0,31}},{P_{0,33}},{P_{3,1}},{P_{3,3}}$ can be obtained.

Submitting $\phi _{,1}^ + ,\phi _{,1}^ + ,{\phi _{,1}},{\phi _{,3}},{P_{0,31}},{P_{0,33}},{P_{3,1}},{P_{3,3}}$ obtained above into (\ref{eq8}), one can obtain,

\begin{equation}\label{eq11}
M\left[ {\begin{array}{*{20}{l}}
A\\
B\\
C\\
D
\end{array}} \right] = \left[ {\begin{array}{*{20}{c}}
{ - \frac{{\chi {B_3}}}{{{\mu _0}{\mu _r}}}E}\\
{ - \frac{{\chi {B_1}}}{{{\mu _0}}}E}\\
0\\
0
\end{array}} \right]
\end{equation}

where, 
\begin{equation*}
M = \left[ {\begin{array}{*{20}{l}}
{i\xi }&{ - [1 + \frac{{2(1 - 2\nu )}}{{G{\mu _0}\mu _r^2}}{\chi ^2}B_3^2]i\xi {\rm{                     }}}&{ - \frac{{\chi {B_3}}}{{{\mu _0}{\mu _r}}}i\xi \left| \xi  \right|}&{(3 - 4\nu )\frac{{\chi {B_3}}}{{{\mu _0}{\mu _r}}}i\xi }\\
{ - \left| \xi  \right|}&{ - {\mu _r}\left| \xi  \right| - \frac{{2(1 - 2\nu )}}{{G{\mu _0}{\mu _r}}}{\chi ^2}{B_1}{B_3}i\xi {\rm{              }}}&{ - \frac{{\chi {B_1}}}{{{\mu _0}}}i\xi \left| \xi  \right|{\rm{ }}}&{(3 - 4\nu )\frac{{\chi {B_1}}}{{{\mu _0}}}i\xi }\\
0&{\frac{\chi }{G}[{B_1}(4\nu  - 1)\left| \xi  \right| + \frac{{{B_3}}}{{{\mu _r}}}(4\nu  - 1)]i\xi {\rm{        }}}&{ - 2i\xi \left| \xi  \right|{\rm{      }}}&{2(1 - 2\nu )i\xi {\rm{   }}}\\
0&{ - \frac{{2\nu }}{G}\chi {B_1}i\xi  - \frac{{\chi {B_3}}}{{G{\mu _r}}}[2(1 - \nu ) + \frac{{\chi  - 2}}{2}]\left| \xi  \right|}&{ - {\xi ^2}{\rm{           }}}&{2(1 - \nu )\left| \xi  \right|{\rm{    }}}
\end{array}} \right]
\end{equation*}

\begin{equation*}
A = A(\xi ),B = B(\xi ),C = C(\xi ),D = D(\xi )\hspace{17em}
\end{equation*}

\begin{equation*}
E = \frac{P}{{2\pi G}}\{ [(1 - 2\nu )\pi  + \pi l\left| \xi  \right|]\cos \beta  + [i\pi \xi l + 2(1 - \nu )\frac{{i\pi \xi }}{{\left| \xi  \right|}}]\sin \beta \} \exp ( - \left| \xi  \right|l)\hspace{10em}
\end{equation*}

From (\ref{eq11}), we can obtain:

\begin{equation}\label{eq12}
A = \frac{{2\chi /{\mu _0}}}{{R_1^2 + R_3^2}}(\frac{{{B_3}{R_1}}}{{\left| \xi  \right|}} + \frac{{{B_3}{R_3}i}}{\xi } - \frac{{{B_1}{R_1}i}}{\xi } + \frac{{{B_1}{R_3}}}{{\left| \xi  \right|}})E
\end{equation}

where, 
\begin{equation*}
\left\{ {\begin{array}{*{20}{l}}
{{R_1} = \frac{{{\chi ^2}{B_1}{B_3}}}{{G{\mu _0}{\mu _r}}}[ - 2 + 6\nu  + 2(1 - \nu ){\mu _r}]}\\
{{R_3} = \frac{{{\chi ^2}B_1^2}}{{G{\mu _0}}}(1 + 2\nu ) + 2({\mu _r} + 1) + \frac{{{\chi ^2}B_3^2}}{{G{\mu _0}\mu _r^2}}[(3 - 4\nu  - 2(1 - \nu ){\mu _r}]}
\end{array}} \right.
\end{equation*}

From (\ref{eq12}) and (\ref{eq9}), we can get,

\begin{equation*}
{\phi ^ + } = \frac{{\chi P}}{{G{\mu _0}(R_1^2 + R_3^2)}}[{B_3}{R_1}({S_1} + {S_2}) + {B_3}{R_3}({S_3} + {S_4}) - {B_1}{R_1}({S_3} + {S_4}) + {B_1}{R_3}({S_1} + {S_2})]
\end{equation*}

where,
\begin{equation*}
\left\{ {\begin{array}{*{20}{l}}
{{S_1} = \{  - (1 - 2\nu )\ln [x_1^2 + {{({x_3} + l)}^2}] + \frac{{l({x_3} + l)}}{{x_1^2 + {{({x_3} + l)}^2}}}\} \cos \beta }\\
{{S_2} = [ - \frac{{l{x_1}}}{{x_1^2 + {{({x_3} + l)}^2}}} + 2(1 - \nu )\arctan \frac{{{x_1}}}{{{x_3} + l}}]\sin \beta }\\
{{S_3} = [ - \frac{{l{x_1}}}{{x_1^2 + {{({x_3} + l)}^2}}} + (1 - 2\nu )\arctan \frac{{{x_1}}}{{{x_3} + l}}]\cos \beta }\\
{{S_4} = \{ 2(1 - \nu )\ln [x_1^2 + {{({x_3} + l)}^2}] - \frac{{l({x_3} + l)}}{{x_1^2 + {{({x_3} + l)}^2}}}\} \sin \beta }
\end{array}} \right.
\end{equation*}

Hence,

\begin{equation}\label{eq13}
{\footnotesize b_3^ +  = {\mu _0}\phi _{,3}^ +  = \frac{{\chi P}}{{G(R_1^2 + R_3^2)}}\left\{ {\begin{array}{*{20}{c}}
{\cos \beta \{ ({B_3}{R_1} + {B_1}{R_3})[ - 2(1 - 2\nu ){Q_1} + {Q_2}] + ({B_3}{R_3} - {B_1}{R_1})[ - (1 - 2\nu ){Q_3} + {Q_4}]\} }\\
{\sin \beta \{ ({B_3}{R_1} + {B_1}{R_3})[{Q_4} - 2(1 - \nu ){Q_3}] + ({B_3}{R_3} - {B_1}{R_1})[4(1 - \nu ){Q_1} - {Q_2}]\} }
\end{array}} \right\}}
\end{equation}

\begin{equation}\label{eq14}
{\footnotesize b_1^ +  = {\mu _0}\phi _{,1}^ +  = \frac{{\chi P}}{{G(R_1^2 + R_3^2)}}\left\{ {\begin{array}{*{20}{c}}
{\cos \beta \{ ({B_3}{R_1} + {B_1}{R_3})[ - 2(1 - 2\nu ){Q_3} - {Q_4}] + ({B_3}{R_3} - {B_1}{R_1})[(1 - 2\nu ){Q_1} + {Q_2}]\} }\\
{\sin \beta \{ ({B_3}{R_1} + {B_1}{R_3})[2(1 - \nu ){Q_1} + {Q_2}] + ({B_3}{R_3} - {B_1}{R_1})[4(1 - \nu ){Q_3} + {Q_4}]\} }
\end{array}} \right\}}
\end{equation}

where,
\begin{equation*}
\left\{ {\begin{array}{*{20}{l}}
{{Q_1} = \frac{{{x_3} + l}}{{x_1^2 + {{({x_3} + l)}^2}}}}&{{Q_2} = \frac{{lx_1^2 - l{{({x_3} + l)}^2}}}{{{{[x_1^2 + {{({x_3} + l)}^2}]}^2}}}}\\
{{Q_3} = \frac{{{x_1}}}{{x_1^2 + {{({x_3} + l)}^2}}}}&{{Q_4} = \frac{{2l{x_1}({x_3} + l)}}{{{{[x_1^2 + {{({x_3} + l)}^2}]}^2}}}}
\end{array}} \right.
\end{equation*}

\begin{figure}[htb]
\setlength{\belowcaptionskip}{0pt}
\centering
\begin{overpic}[scale=0.5]{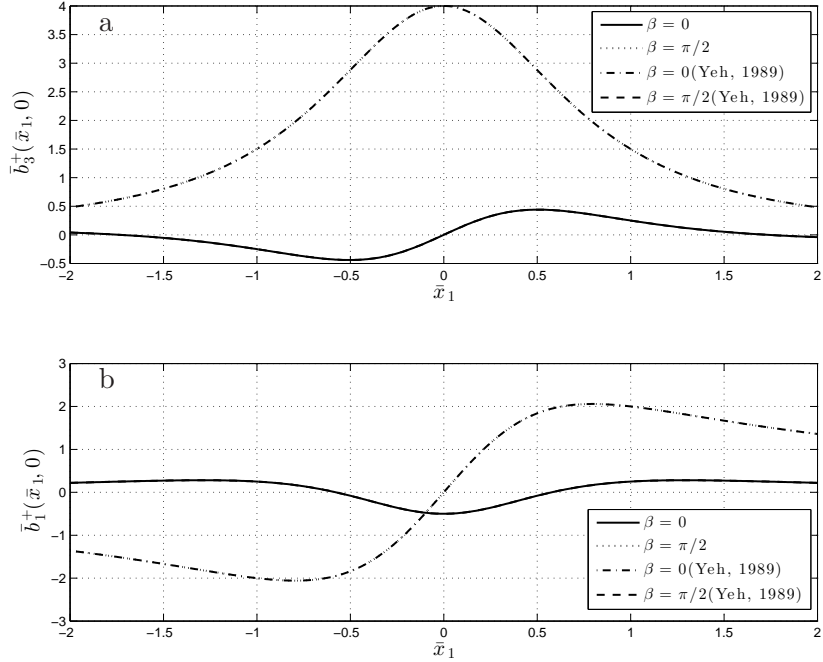}
%\begin{overpic}[scale=0.5,grid,tics=4]{Figure2.eps}
\put(16,33){b}
\put(16,70){a}
\end{overpic}
\renewcommand{\figurename}{Fig.}
\caption{The variation of the normalized induced magnetic fields $\bar b_i^ +  = {b_i^ + G{R_3}l}/({\chi P{B_3}})$  $(i = 1, 3)$ at $x_3=0$  with normalized ${\bar x_1} = {x_1}/l$  , when $\theta=\pi/2$  and the $\beta=0,\pi/2$  respectively.}
\label{fig2}
\end{figure}

To check our results, Figure \ref{fig2} shows the variation of the normalized induced magnetic fields $\bar b_i^ +$ ($\bar b_3^ +$ in Figure \ref{fig2}a; $\bar b_1^ +$ in Figure \ref{fig2}b; $\bar b_i^ +$ is defined below)  at $x_3=0$ with normalized ${\bar x_1} = {x_1}/l$ , when ${\bf{B}} = {B_3}{{\bf{e}}_3}$ and the ${\bf{f}} = {f_3}{{\bf{e}}_3}$ and/or ${\bf{f}} = {f_1}{{\bf{e}}_1}$. Here $\chi  = {10^{ - 4}}$, $\nu  = 0.2$ , ${\mu _0}{\rm{ = 4.0}} \times {\rm{1}}{{\rm{0}}^{ - 7}}{\rm{N}}{{\rm{A}}^{ - 2}}$ , ${\mu _r} \approx {\rm{1}}$, which are from the granite -- the representation rock for the crust. The averaged ${\bf{B}}$ from the earth’s main field is about $50\mu {\rm{T}}$. Hence, ${R_1} \to 0$ , and $\bar b_i^ +  = \frac{{b_i^ + G{R_3}l}}{{\chi P{B_3}}}$. It can be seen from Figure \ref{fig2} that our results for these cases are the same as those from Yeh (1989).  It can also be seen that $\bar b_3^ + $ passes through zero and $\bar b_1^ + $ reaches its extremum when ${\bar x_1} = 0$, ${\bf{B}} = {B_3}{{\bf{e}}_3}$ and ${\bf{f}} = {f_1}{{\bf{e}}_1}$, which is in accord with the phenomena observed in non-destructive test (eg., Doubov, 1997). 

\begin{figure}[htb]
\setlength{\belowcaptionskip}{0pt}
\centering
\begin{overpic}[scale=0.32]{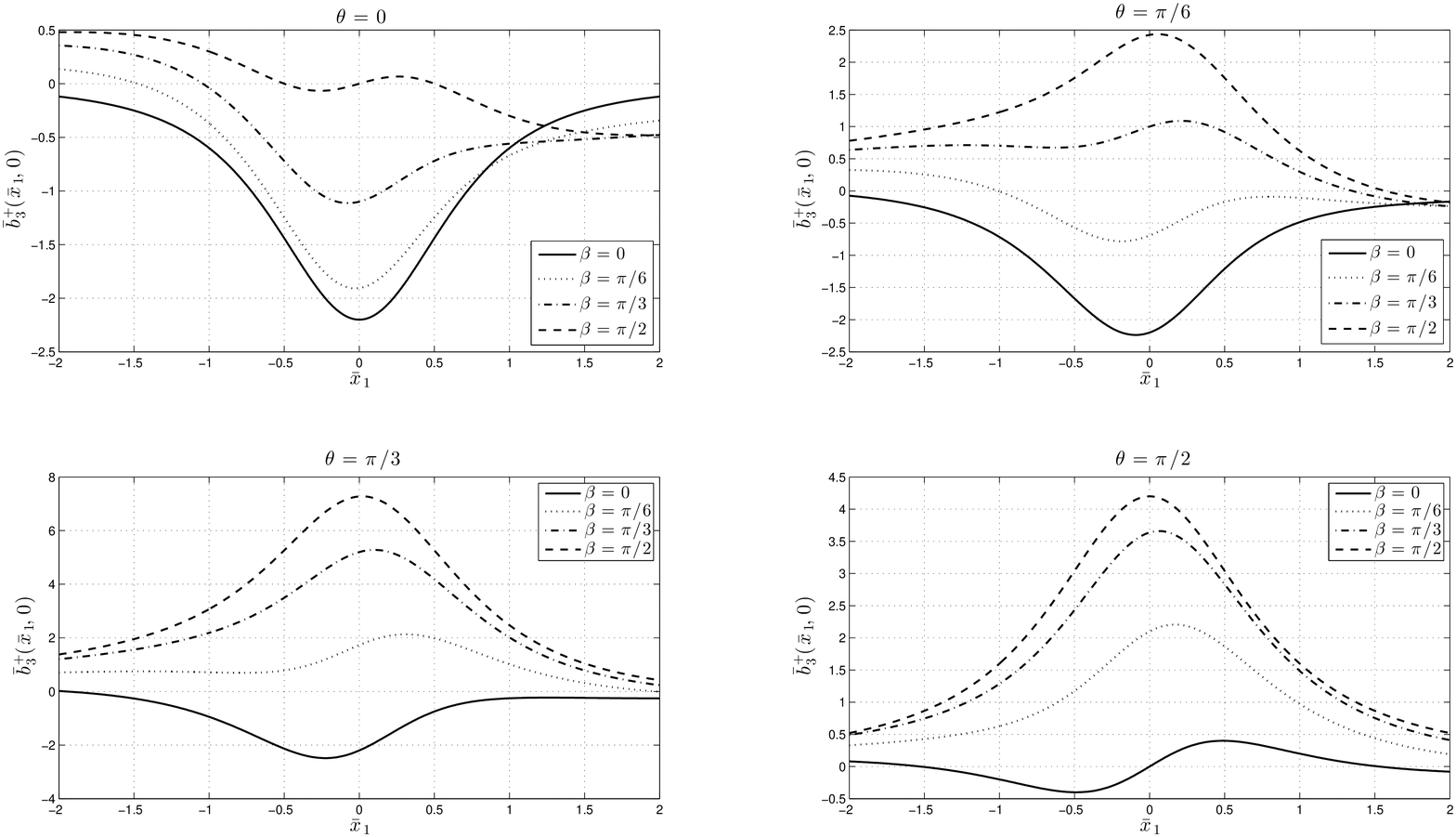}
\end{overpic}
\renewcommand{\figurename}{Fig.}
\caption{The variation of the normalized induced magnetic fields $\bar b_3^ + $  at $x_3=0$  with normalized ${\bar x_1} = {x_1}/l$, when $\theta  = 0,{\rm{ }}\pi /6,{\rm{ }}\pi /3,{\rm{ }}\pi /2$ and $\beta  = 0,{\rm{ }}\pi /6,{\rm{ }}\pi /3,{\rm{ }}\pi /2$, respectively.}
\label{fig3}
\end{figure}

\begin{figure}[htb]
\setlength{\belowcaptionskip}{0pt}
\centering
\begin{overpic}[scale=0.32]{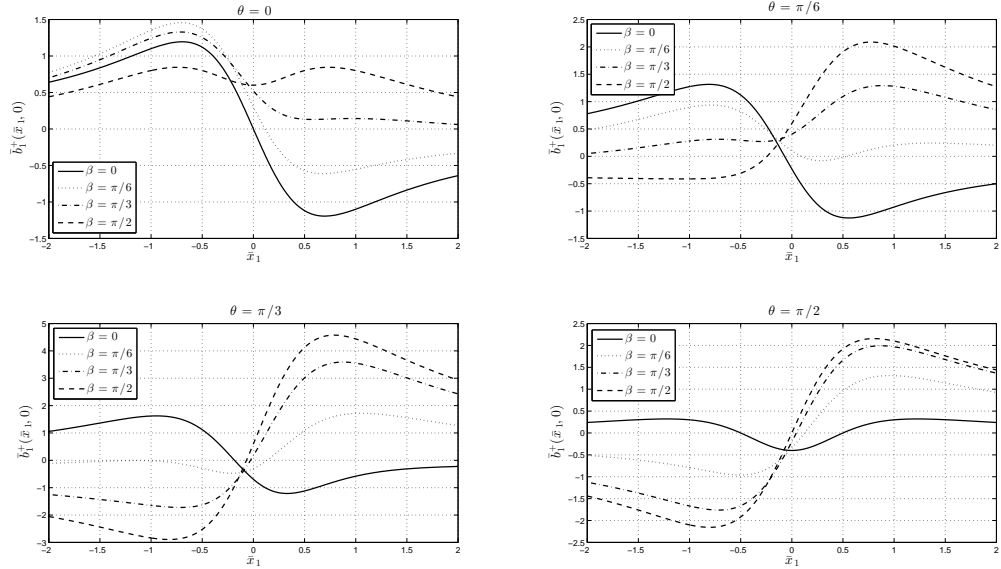}
\end{overpic}
\renewcommand{\figurename}{Fig.}
\caption{The variation of the normalized induced magnetic fields $\bar b_1^ + $  at $x_3=0$  with normalized ${\bar x_1} = {x_1}/l$, when $\theta  = 0,{\rm{ }}\pi /6,{\rm{ }}\pi /3,{\rm{ }}\pi /2$ and $\beta  = 0,{\rm{ }}\pi /6,{\rm{ }}\pi /3,{\rm{ }}\pi /2$, respectively.}
\label{fig4}
\end{figure}

Figure \ref{fig3} and Figure \ref{fig4} show the variation of the normalized induced magnetic fields $\bar b_3^ +$ and $\bar b_1^ +$ at $x_3=0$ with normalized $\bar{x}_1$, respectively, when ${\bf{B}} = {B_1}{{\bf{e}}_1} + {B_3}{{\bf{e}}_3}$ and the ${\bf{f}} = {f_1}{{\bf{e}}_1} + {f_3}{{\bf{e}}_3}$. Only the cases that $\theta  = \arctan (\frac{{{B_3}}}{{{B_1}}}) = 0,{\rm{ }}\frac{\pi }{6},{\rm{ }}\frac{\pi }{3},{\rm{ }}\frac{\pi }{2}$ and $\beta  = \arctan (\frac{{{f_3}}}{{{f_1}}}) = 0,{\rm{ }}\frac{\pi }{6},{\rm{ }}\frac{\pi }{3},{\rm{ }}\frac{\pi }{2}$ are shown.  It can be seen that the distributions of $\bar b_1^ +$ at ${\bar x_1} = 0$ are too complicated to be differentiated, while those of $\bar b_3^ + $
are relative simpler. Some distributions of $\bar b_3^ + $ at ${\bar x_1} = 0$ will be used in the following.

\section{Application of the induced magnetic field $\bar b_3^ + $ at ${\bar x_1} = 0$ by a line mechanical singularity to the seismicities in the continental crust}\label{sec4}

The recent earthquakes, e.g., 2010 Haiti and Chile, 2011 Tohoku in Japan, again painfully reminded us of the destructive impact of shallow seismicities and the importance of the available and reliable earthquake risk information. At present the risk information of seismicity is mainly from instrumental and historical catalogs (e.g., Giardini et al. 1999；Kagan and Jackson, 2011), geodesy or plate tectonics (e.g., Bird et al. 2010). In the Seismic Hazard by the Global Earthquake Model project (http://www.globalquakemodel.org), data of historical Earthquake Catalogue and Archive (1000-1903), instrumental Earthquake Catalogue (1900-2009), active faults , geodetic Strain Rate, and ground motion, are included. Such information is useful and helpful to gain a better understanding of the distribution of the seismicities in the continental crust, and then to minimize loss of life, property damage and social and economic disruption due to earthquakes. More data or constraints from other geo-science are needed in order to get the risk information of seismicity above more available and reliable. 

An earthquake can be approximated to a line concentrated force, although this simplification is rough. If so, some results in the following from section \ref{sec3} can be used to constrain the distribution of the epicenters of the earthquakes occurs in the crust. As can be seen from Figure \ref{fig3}, the distributions of $\bar b_3^ + $ are relative simpler, and $\bar b_3^ + (0,0) = 0$  when $\beta  = 0,{\rm{ }}{B_1} = 0$,$\beta  = \pi /2,{\rm{ }}{B_3} = 0$, or $\cos \beta \{ {B_1}[2(1 - 2\nu ) + 1]\}  = \sin \beta \{ {B_3}[4(1 - \nu ) + 1]\}$ (if $\nu  = 0.25$, this means that ${B_1}\cos \beta  = 2{B_3}\sin \beta$). In other words, if  $\bar b_3^ + (0,0) \to 0$  when $\beta$ is small and $B_3$ is dominant, or if $\beta  \to \pi /2$ and $B_1$ is dominant, or $\beta  \to {\tan ^{ - 1}}({B_1}/2{B_3})$ (assuming $\nu=0.25$ ). Hence $\bar b_3^ + (0,0) \to 0$ can be used as a sign for the concentrated force (earthquake) when the direction of the magnetic field is known.

\begin{figure}[htb]
\setlength{\belowcaptionskip}{0pt}
\centering
\begin{overpic}[scale=0.8]{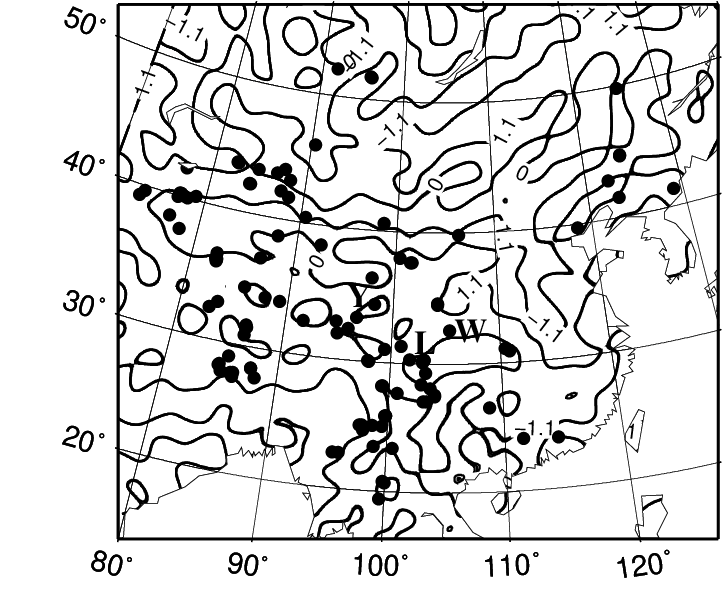}
\end{overpic}
\renewcommand{\figurename}{Fig.}
\caption{The relationship between the zero-contours of the vertical component of the crustal magnetic induction and the distribution of the epicenters for the earthquakes within the crust under the Chinese continent and its adjacent areas($15-55^\circ\mbox{N}$,$70-146^\circ\mbox{E}$). The filled circle are the earthquakes whose magnitude (Ms) are larger than 5.0. The curves labeled with $0$ are zero-contours of the vertical component of the crustal magnetic induction at the surface, and those labeled with $+1.1$ or $-1.1$ are the RMS contours. The earthquakes occurred after 2010, which are from Chinese Seismic Networks (CSN). Zero-contours of the vertical component of the crustal magnetic induction at the surface are from MF7.}
\label{fig5}
\end{figure}

Figure \ref{fig5} shows the distribution of the epicenter of the earthquakes occurred in the crust of the Chinese continent after 2010, along with the zero-contours of the $\bar b_3^ + (0,0)$ from the crustal magnetic field model MF7 which have an RMS of 1.1 nT (Maus et al. 2008). The seismic events are from Chinese Seismic Networks (CSN). The lower bound of the seismic magnitude is 5.0 (Ms), for earthquakes whose magnitude less than 5.0 are less dangerous to us and have less accuracy in their locations.  According to the depth of the crust from a global crustal model of CRUST 1.0 (Laske et al. 2013), those seismic events within the crust are selected out.

MF7 is used here for two reasons: (1) Only satellite data is used in MF7, so that the relative results can be extrapolated to those regions without the marine, aeromagnetic and ground magnetic surveys but only with satellite data, although there are other crustal magnetic field models, for example, NGDC-720 (http://geomag.org/models/ngdc720.html). In fact, magnetic models similar to NGDC-720 should be used in smaller regions because they includes more detailed information than MF7 does. (2) MF7 has minimized some other irrelative information, for example, the time-varying contributions due to the interaction between the Earth's core field and the solar wind, which is usually included in the original satellite data observed.

It can be seen from Figure \ref{fig5} that most of the epicenters of the seismic events are on the zero-contours of the $\bar b_3^ + (0,0)$, which account approximately for 65\% of the seismic events; Some epicenters are near the zero-contours of the $\bar b_3^ + (0,0)$, which account approximately for 22\% of the seismic events; A few epicenters are away from the zero-contours of the $\bar b_3^ + (0,0)$, which account approximately for 13\% of the seismic events. The Lushan earthquake, which is marked by “L”, is on the zero-contours of the $\bar b_3^ + (0,0)$. In summary, the epicenters are on or near the zero-contours of the $\bar b_3^ + (0,0)$ account approximately for 80\% of the total seismic events in the Figure \ref{fig5}, thus $\bar b_3^ + (0,0)\to 0$ can be used as a sign for the concentrated force (earthquake) when the direction of the magnetic field is known.

It can also be seen from Figure 5 that the earthquake of Wenchuan and Yushu, which are marked by “W” and “Y” and occurred in 2008 and 2010, respectively, are also on the zero-contours of the $\bar b_3^ + (0,0)$, for our model calculated is an elastic one.

\begin{figure}[htb]
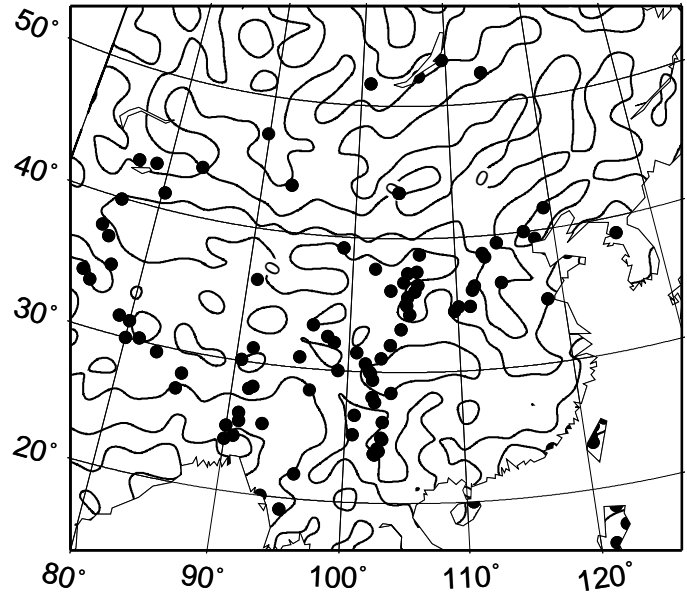

\setlength{\belowcaptionskip}{0pt}
\centering
\begin{overpic}[scale=0.8]{Figure6.eps}
\end{overpic}
\renewcommand{\figurename}{Fig.}
\caption{The relationship between the zero-contours of the vertical component of the crustal magnetic induction and the distribution of the epicenters for the historical earthquakes within the crust under the Chinese continent and its adjacent areas ($15-55^\circ\mbox{N}$, $70-146^\circ\mbox{E}$). The filled circles are the historical earthquakes. The curves labeled with '0' are zero-contours of the vertical component of the crustal magnetic induction at the surface. The earthquakes are from GEM project and Sha et al. (2009). Zero-contours of the vertical component of the crustal magnetic induction at the surface are from MF7.}
\label{fig6}
\end{figure}

Figure \ref{fig6} shows the relationship between the zero-contours of the vertical component of the crustal magnetic induction and the distribution of the epicenters for the historical earthquakes within the crust under the Chinese continent and its adjacent areas. The data of historical earthquakes are from GEM project and Sha et al. (2009). It can be seen that most of the epicenters of the seismic events are also on the zero-contours of the $\bar b_3^ + (0,0)$, which account approximately for 37\% of the seismic events; Some epicenters are near the zero-contours of the $\bar b_3^ + (0,0)$, which account approximately for 43\% of the seismic events; A few epicenters are away from the zero-contours of the $\bar b_3^ + (0,0)$, which account approximately for 20\% of the seismic events. In summary, the epicenters are on or near the zero-contours of the $\bar b_3^ + (0,0)$ account approximately for 80\% of the total seismic events in the Figure \ref{fig6}, thus  $\bar b_3^ + (0,0)\to 0$ can also be used as an auxiliary sign in the determination of the great historical earthquake.

\section{Discussions}

The physical fields of the earth are multi-physical coupled fields. Here we studied a simple coupled field in which the effects of magnetic field and elastic deformation field are taken into account. Equation (\ref{eq13}) and (\ref{eq14}) show that the line concentrated force will induced a perturbed magnetic field. This may be resulted from the rearrangement of the magnetic sources because of the elastic deformation induced by the concentrated force. It can also be seen that the induced magnetic field will indicate the line concentrated force in reverse, and this can be used to constrain the distribution of the line concentrated force, for example, an earthquake in the crust.

Seismicities in the continental crust are more dangerous, but they distribute diffusely. It is an interesting problem to study the relationship between the earthquake and other geo-science data. Some studies has been carried out on the relationship between the satellite magnetic field and earthquakes (e.g., Li et al. 1992, Ding et al., 2011), or on assessing intraplate earthquake hazards by satellite magnetic data (e.g., Von Frese1 et al., 2008; Taylor et al., 2008). The relative results show that satellite magnetic anomaly has a close relationship with the distribution of the earthquakes; and it is possible to constrain the epicenter distribution by the satellite magnetic field as well. However, the relationship obtained by these studies is too complicated to be employed in a convenient way. For example, that of Li et al. (1992) showed that earthquakes distributed along the geo-magnetic anomalies or their gradient zone, or their corners, or their junction points, which have some problems in convenience and availability. In this paper, a simpler flag, i.e., the zero contours of the $\bar b_3^ + (0,0)$ is used to constraint to the shallow seismicities in the continental crust.

Results from Figure \ref{fig5} show that there is a close relationship between the epicenters and the zero-contours of the $\bar b_3^ + (0,0)$ observed by the satellite. It can be deduced that the regions where are at these zero-contours, especially those in active tectonic zones, could be the possible seismogenic zones in the future. Consequently, these zero-contours of the $\bar b_3^ + (0,0)$ could be used as a constraint to the risk information for the continental shallow seismicities, although this constraint could not provide any information on when the earthquake occur, which should be aided with other geo-science data and techniques, because our model is only an elastic one and it is unknown when and how the concentrated forces come into being.

Simultaneously, it can be seen from Figure \ref{fig5} that some epicenters are not at the zero-contours of the $\bar b_3^ + (0,0)$, which are results from many reasons. Firstly if the conditions in section \ref{sec3} are not satisfied, a non-zero $\bar b_3^ + (0,0)$ could be induced, as shown in Figure \ref{fig2} and Figure \ref{fig3}. However, it can be seen that these non-zero $\bar b_3^ + (0,0)$ are near zero, except that $P$ is very large, or $l$ is very small.  We take the case in Figure \ref{fig2} as an example. If the parameters, such as $\chi$, in section \ref{sec3} are used; Assuming $l$ is 10 km and $P$ is $10^{15} \mbox{Jm}^{-2}$, which is corresponding to that an earthquake of Ms=5.0 occurs on an area of $1.0 \mbox{m}^2$, although there is no agreement on how much $P$ is at present. Such that $\frac{{\chi P{B_3}}}{{G{R_3}l}} \approx 6{\rm{ nT}}$. Therefore, if the conditions in section \ref{sec3} for $\bar b_3^ + (0,0)$ does not hold, the corresponding deviation is small, except that $P$ is very large and/or  $l$ is very small. Secondly these small non-zero $\bar b_3^ + (0,0)$  may contribute to the uncertainty of earthquake location. Thirdly, the non-zero $\bar b_3^ + (0,0)$ may come from the satellite geo-magnetic field used here, because it is an only a model, and there may be no enough or even no data in some areas. 

In fact, the magnetic induction above the half-space consists of two parts: one is that induced by the concentrated force, the other is the applied magnetic field.  Therefore, it is very important to separate completely and accurately the magnetic field induced by the concentrated forces from the earth's magnetic field observed. It seems that there is no better method to achieve this at present.

Here MF7 is resolved from spherical harmonic degree 16 to 133, for a clean separation of the lithospheric signal from ionospheric and magnetospheric noise sources could be achieved (Maus et al., 2008). However, this maybe eliminates the long-wave length part of the induced field. And the abnormal field from the local crustal rocks, or remanent magnetism from the crustal rocks, may also still be included as well. They may have influences on or complicate the zero-contours of $\bar b_3^ + (0,0)$ of the MF7.

Despite the above, it can be seen from Figure \ref{fig5} that the zero-contours of the $\bar b_3^ + (0,0)$ from satellite could be as a geophysical constraint to the shallow seismicities in the continental crust, which are independent on risk information from geodesy and tectonics. They can also be used as an early and rough monitor for the shallow seismicities before the considerable work from instrument, geodesy and tectonics. With the accumulation of data and advances in processing data, this constraint to the risk information for earthquake will be more available and reliable.

The seismic events used in this paper occurred between 2011-2014; while the MF7 used the satellite data observed between 2008 and 2010. It can be seen that our results have some ability of "prewarning", although our model is an elastic one, and it is unknown when and how the concentrated forces come into being.

\section{Conclusions}

After the analysis and discussions above, the following conclusions can be drawn,

It is helpful to study the multi-physic fields and their coupled effects for understanding the properties of the complicated system of the earth’s. In this paper the interaction between the magnetic field and elastic deformation field in the crust is studied in a simplified way. The results show that the line concentrated force will induced a perturbed magnetic field; The induced magnetic field will indicate the line concentrated force in reverse.  

   For a line concentrated force in a half space which is magnetized uniformly vertically (or horizontally), theoretical calculation by us shows that the vertical component of the magnetic induction induced by the concentrated force is zero at the origin, when the applied magnetic field is perpendicular to the line concentrated force, or the applied magnetic field and the line concentrated force satisfy some conditions. 

  Most of the epicenters of the earthquakes in the continental crust are at or near the zero-contours of the $\bar b_3^ + (0,0)$  observed from satellite. In the Chinese continental crust, these epicenters account for more than 80\% epicenters of the earthquakes and/or great historical earthquakes. These regions, especially those in active tectonic zones, could be the possible seismogenic zones in the future.

   The zero-contours of the $\bar b_3^ + (0,0)$ from satellite could be as geophysical constraints to the risk information on the shallow seismicities; Or they can be used as an early monitors for the shallow seismicities in the continental crust, or an auxiliary sign in the determination of the great historical earthquake. These constraints or monitors are more available and helpful.

\vspace{5em}

{\Large\bf Acknowledgment}

This work is supported by the National Natural Science Foundation of China (Grant No. 41074064, 40604006).

\vspace{10em}

\def\thebibliography#1{
%\centerline
{\Large\bf  References}\list
% The \arabic{enumi} command has been removed because we don't use
% numbered citations in astronomical journals.
 {}{\setlength\labelwidth{1.4em}\leftmargin\labelwidth
 \setlength\parsep{0pt}\setlength\itemsep{.3\baselineskip}
 \setlength{\itemindent}{-\leftmargin}
 \usecounter{enumi}}
 \def\newblock{\hskip .11em plus .33em minus -.07em}
 \sloppy
 \sfcode`\.=1000\relax}
\let\endthebibliography=\endlist


\begin{thebibliography}{99}
\bibitem{}
Bird P, Kreemer C, Holt WE. A Long-term Forecast of Shallow Seismicity Based on the Global Strain Rate Map. Seism. Res. Lett, 2010, 81(2):184194. doi:10.1785/gssrl.81.2.184
\bibitem{}
Brown W F Jr, Magnetoelastic interactions, Springer-Verlag, New York, 1966
\bibitem{}
Ding J H, Lu Z Y, Yu S R. Introduction of seismogeomagnetism. Hefei: Press of University of Science and Technology of China, 2011
\bibitem{}
Dorfmann A L, Ogden R W, Nonlinear Theory of Electroelastic and Magnetoelastic,  Interactions, Springer, New York, 2014
Doubov A A. A study of metal properties using the method of magnetic memory.  Metal Science and Heat Treatment, 1997, 39(9-10): 401-402
\bibitem{}
Giardini D, Grunthal G, Shedlock K, et al. Global Seismic Hazard Map, 1:35,000,000. Global Seismic Hazard Assessment Program, UN/International Decade of Natural Disaster Reduction, International Lithosphere,1999
\bibitem{}
Huang K F, Wang M Z, Complete solution of the linear magneto elasticity and the magnetic fields in a magnetized elastic half-space, J Appl. Mech., AMSE, 1995,  62(4), 930-934
\bibitem{}
Kagan Y Y,  Jackson D D,  Global earthquake forecasts, Geophys. J. Int., 2011, 184(2): 759-776, doi: 10.1111/j.1365-246X.2010.04857.x.
\bibitem{}
Laske G, Masters G, Ma Z,  Pasyanos M, Update on CRUST1.0 - A 1-degree Global Model of Earth's Crust, Geophys. Res. , 2013, Abstracts, 15, Abstract EGU2013-2658
\bibitem{}
Li J F, Wu Y Z, An Z C, et al. The relationship between magnetic anomaly and the distribution of earthquakes in China continent. South China Journal of Seismology,1992, 12(1):1116
\bibitem{}
Lin C B, Yeh C S, The magneto elastic problem of a crack in a soft ferromagnetic solid, Int. J. Solids Struct., 2002, 39: 1-17
\bibitem{}
Maugin G A, A continuum approach to magnonphonon couplings-I: General equations, background solution, Int. J. Eng. Sci, 1979, 17 (10): 1073-1091
\bibitem{}
Maus S, Yin F,Lühr H, et al. Resolution of direction of oceanic magnetic lineations by the sixth-generation lithospheric magnetic field model from CHAMP satellite magnetic measurements. Geochem. Geophys. Geosyst.,2008, 9: Q07021. doi:10.1029/2008GC001949
\bibitem{}
Pao Y H and Yeh C S,  A linear theory for soft ferromagnetic elastics solids, Int. J. Eng. Sci., 1973, 11, 415-436
\bibitem{}
Qin F, Yan D M, Zhang X F, Perturbed magnetic fields generated by deformation of structures in earth magnetic field. Acta Mechanic Sinica , 2006, 38(6): 799–806  (in Chinese)
\bibitem{}
Sha H J, Liu Y W, Liu D Y, Preliminary study on adjusting effect of great earthquake in Chinese mainline, Earthquake research in China, 2009, 25 (2): 193-198 (in Chinese)
\bibitem{}
Shindo Y. The linear magneto elastic problem for a soft ferromagnetic elastic solid with a finite crack, J Appl. Mech., AMSE, 1977, 44(1), 47-50
\bibitem{}
Taylor PT, Kim RH, Kutina J, Johnson GL. Geohazard assessment from satellite magnetic data modeling with examples from the Arctic Margin along the Canada Basin and the Korean Peninsula along 40o N (latitude) parallel. Earth Planets Space, 2008, 60: 497503
\bibitem{}
Tiersten H F, Coupled magneto mechanical equations for magnetically saturated insulators, J. Math. phys., 1964,5, 1298-1318
\bibitem{}
Von Frese1 RRB, Kim JW, Hernandez O, et al. Assessing intraplate earthquake hazards from satellite geopotential field observations. Earth Planets Space, 2008, 60: 487495
\bibitem{}
Wang M Z, Advanced Elastics, Peking University Press, Beijing, 2002
\bibitem{}
Wei R Q, Yu L, Constrains from satellite crustal magnetic field on the distribution of the epicenters for earthquakes in the continental crust of China, Chinese J. Geophys. (in Chinese), 2012, 55 (8): 2643-2650, doi: 10. 6038/j. issn. 0001-5733. 2012.08.016
\bibitem{}
Yeh C S, Magnetic fields generated by a mechanical singularity in a magnetized elastic half-plane, J. Appl. Mech., AMSE,  1989, 56: 89-95
\bibitem{}
Yeh C S, Magnetic fields generated by a tension fault. Bulletin of the College of Engineering, National Taiwan University , 1987, 40, 47–56
\bibitem{}
Yeh C S, Ren C W, Magnetic fields generated by a mechanical singularity in a magnetized anisotropic half plane,  J. Mech., 2010, 26(2): 173-186
\bibitem{}
Zhou Y H, Zheng X J, A theoretical model of magneto elastic buckling for soft ferromagnetic thin plates, Acta Mechanica Sinica, 1996, 12 (3): 213-224

\end{thebibliography}
\end{document}